\documentclass[twocolumn,showpacs,preprintnumbers,amsmath,amssymb,prl,aps]{revtex4}

\usepackage{graphicx}% Include figure files
\usepackage{dcolumn}% Align table columns on decimal point
\usepackage{bm}% bold math
\begin{document}

\title{Surface bulk differences in a conventional superconductor, ZrB$_{12}$}

\author{Sangeeta Thakur,$^1$ Deepnarayan Biswas,$^1$ Nishaina
Sahadev,$^1$ P. K. Biswas,$^2$\footnote{Present address: Paul
Scherrer Institut, CH-5232 Villigen PSI, SWITZERLAND} G.
Balakrishnan,$^2$ and Kalobaran Maiti$^1$}

\altaffiliation{Corresponding author: kbmaiti@tifr.res.in}

\affiliation{$^1$ Department of Condense Matter Physics and Materials' Science,
Tata Institute of Fundamental Research, Colaba, Mumbai-400005, India\\
$^2$ Department of Physics, University of Warwick, Coventry, CV4
7AL, UK}

\date{\today}

\begin{abstract}
We studied the electronic structure of a conventional
superconductor, ZrB$_{12}$ using high resolution $x$-photoemission
spectroscopy and single crystalline samples. Experimental results
with different bulk sensitivity reveals boron deficiency and
different valence states of Zr at the surface relative to the bulk.
Signature of a satellite features is observed in the Zr core level
spectra corresponding to the bulk of the material suggesting
importance of electron correlation among the conduction electrons in
the bulk while the surface appears to be uncorrelated. These results
provide an insight in fabricating devices based on such
superconductors.
\end{abstract}

%\pacs{74.70.-b, 79.60.Bm, 73.20.At}

\maketitle

%\maketitle
%\section{Introduction}

Superconductors play an important role in technological applications
including various medical tools and are potential candidates for
future applications such as lossless power transmission, maglev
trains etc. Fabrication of devices requires good knowledge of the
surface and bulk properties, which are often found to be different
in many systems.\cite{surface} Here, we considered a conventional
superconductor, ZrB$_{12}$; the electronic properties of these
systems are well captured by the Bardeen-Cooper-Schriefer (BCS)
theory and comes among the simplest cases of the superconducting
materials.

ZrB$_{12}$ exhibits highest superconducting transition temperature
($T_c \sim$~6~K) in $M$B$_{12}$ family \cite{Matthias}. In the
crystal, Zr atoms are surrounded by 24 boron atoms arranged on a
truncated octahedron and has the smallest lattice constant (space
group\cite{Glaser,Paderno} $Fm3m$, $a$ = 7.4075 \AA) of all known
dodecaborides.\cite{Fisk} Scanning tunneling spectroscopy and
magnetization measurements show that ZrB$_{12}$ single crystal has
excellent surface properties.\cite{Tsindlekht} Enthalpy
measurements\cite{Stout} in Zr$_{0.6}$Y$_{0.4}$B$_{12}$ suggest the
most probable valence state of Zr to be (+4). In contrast, the
$x$-ray photoemission spectroscopic measurements on polycrystalline
ZrB$_{12}$ sample indicates neutral state of Zr; the earlier
observation of (+4) valence state is attributed to the impurity
phase. Clearly, the microscopic details of the electronic properties
remain to be puzzling and require high resolution studies as
observed in other complex systems.\cite{Bindu} We employed high
resolution $x$-photoemission spectroscopy to investigate the
electronic structure of single crystalline ZrB$_{12}$. $X$-rays of
varied photon energies in the photoemission spectroscopy reveal
interesting differences in the bulk and surface electronic
structures.

%\section{Experimental}

Single crystals of ZrB$_{12}$ were grown by floating zone technique
as described elsewhere.\cite{Geetha} The photoemission measurements
were performed using a Gammadata Scienta R4000 WAL analyzer and
monochromatic Al $K\alpha$ radiations ($h\nu_1$ = 1486.6 eV) with
the energy resolution set to 400 meV. Hard $x$-ray photoemission
(XP) measurements were carried out at P09 beam line, PETRA III
Hamburg, Germany with 5947.9 eV ($h\nu_2$) photon energy and an
electron analyzer from Specs GmbH with the energy resolution set to
150 meV. The melt grown single crystals were hard and have no
cleavage plane. Therefore, we followed surface cleaning procedures
such as fracturing using a post on the sample as well as scraping
using a diamond file in the vacuum chamber with vacuum better than
3$\times$10$^{-11}$ torr. After sample surface preparation in the
ultra-high vacuum chamber, it was transported to the experiment
chamber without exposing to atmosphere so that the surface remains
clean for the photoemission measurements. The angle integrated
spectra were found to be reproducible after the surface cleaning
cycles and no signature of impurity was found in the spectra. The
temperature variation down to 10 K was achieved by an open cycle He
cryostat from Advanced Research systems, USA.

%\section{Results and discussion}

The Zr 3$d$ core level spectra were probed using 1486.6 eV (=
$h\nu_1$) and 5947.9 eV (= $h\nu_2$) photon energies. The
photoelectron escape depth, $\lambda$ (= the distance traveled by
the photoelectrons without inelastic scattering) can be varied by
varying the incident photon energy.\cite{surface}  The value of
$\lambda$ of the conduction electrons is close to 20 \AA\ for
$h\nu_1$ and about 40 \AA\ for $h\nu_2$. Assuming
$\lambda\propto\sqrt{KE}$ at higher kinetic energies (KE), $\lambda$
for Zr 3$d$ electrons would be 18.7 \AA\ and 39.5 \AA\ for $h\nu_1$
and $h\nu_2$, respectively. Thus, at $h\nu_2$, the photoemission
spectrum essentially represents the bulk electronic structure of the
sample.

\begin{figure}
\vspace{-4ex}
 \includegraphics[scale=0.4]{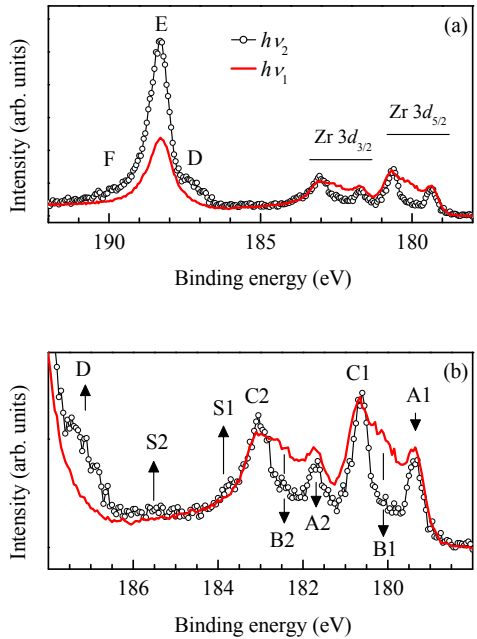}
 \vspace{-12ex} \caption{(a) Zr 3$d$ core level spectra of ZrB$_{12}$
using 1486.6 eV ($h\nu_1$) and 5947.9 eV ($h\nu_2$) photon energies.
Photoemission spectra exhibit distinct signature of satellites. (b)
Zr 3$d$ spectral region is shown with enhanced $y$-scale.}
 \vspace{-2ex}
\end{figure}

The Zr 3$d$ and B 1$s$ core level spectra collected using $h\nu_1$
and $h\nu_2$ are shown in Fig. 1. At $h\nu_1$, the B 1$s$ feature
exhibit a sharp asymmetric shape as expected in a metallic system.
The intensity of the B 1$s$ feature relative to the integrated Zr
4$d$ signal increases significantly at $h\nu_2$ along with an
additional feature, D around 187 eV binding energy and increased
intensity at higher binding energy side (feature, F). Substantial
increase in intensity at the larger probing depth suggests B
deficiency at the surface. This effect is found to be independent of
the surface preparation consistent with earlier results on single
crystal surface cut by diamond saw without chemical etching of the
damaged surface layer.\cite{Lortz} Such boron deficiency at the
surface can arise due to the poorer sticking of borons at the top of
the huge B$_{12}$ cages at the surface.

The Zr 3$d$ spectral region (179 - 186 eV) exhibits multiple
distinct features denoted by A1, A2, B1, B2, C1 and C2 in the figure
- 1's and 2's are used for 3$d_{5/2}$ and 3$d_{3/2}$ signals,
respectively. The spin-orbit splitting is found to be about 2.3 eV.
The intensities of the features changes significantly with the
change in surface sensitivity of the technique indicating
significantly different surface and bulk electronic structure. At
$h\nu_2$, the intensity of the features, B1 and B2 become almost
insignificant although it has large intensity in the $h\nu_1$
spectrum indicating surface nature of these peaks. The features, A1
and A2 reduces in intensity in the bulk sensitive $h\nu_2$ spectrum.
Evidently, the bulk electronic structure is dominated by the
contributions from the features C1 and C2. Two additional features
S1 and S2 are also observed in the spectra - the intensity of these
features become stronger in the bulk sensitive $h\nu_2$ spectrum
again suggesting their bulk nature.

\begin{figure}
\vspace{-4ex}
 \includegraphics[scale=0.4]{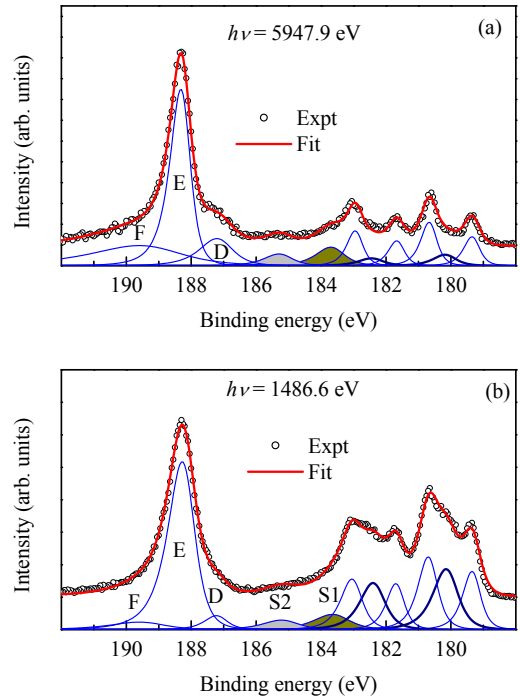}
 \vspace{-8ex}
 \caption{Fits of the core level spectra collected using (a) 1486.6 eV and
(b) 5947.9 eV photon energies. The component peaks are shown by
solid lines.}
 \vspace{-2ex}
\end{figure}

To determine the energy position and relative weight of the spectral
features, we have fit the experimental spectra with a set of
asymmetric Doniach-\v{S}unji\'{c}-type lineshapes,\cite{Doniach}
which is expected in metallic samples due to low energy excitations
across $\epsilon_F$ in the photoemission final states. Three
distinct features in each of the spin-orbit split signals were
represented by three peaks as shown in Fig. 2 and the spin-orbit
splitting is set at the experimental value of 2.3 eV. The least
square error fitting method provides the intensity ratio of the spin
orbit split features close to their multiplicity.

The thick solid lines superimposed over the experimental data in
Fig. 2 exhibit good representation of the experimental spectra. The
binding energies of the features A1 and C1 are found to be 179.4 eV
and 180.7 eV - these energies correspond well with the 3$d$ peak
positions observed for Zr$^{0}$ and Zr$^{2+}$ species, respectively.
This is different from the spectra observed from polycrystalline
samples containing signature of boron oxides.\cite{Flores} Such
multiple valency of Zr have been observed in Zr halides,
oxides,\cite{Morant,Wells,Sarma} and other dodecaborides,
YbB$_{12}$, UB$_{12}$, {\it etc.} as well.\cite{Stout,Takahashi} The
binding energy of feature B1 is found to be about 180.2 eV, which
corresponds to Zr$^+$.

Evidently, the results on ZrB$_{12}$ reveal unusual mixed valency of
Zr - different kinds of mixed valency at the surface and bulk of the
sample. The $h\nu_1$ spectrum exhibits distinct signature of the
feature 'B's corresponding to Zr$^+$ species (thick solid lines in
the figure) and a larger intensities of the features 'A's relative
to that of 'C's. This suggests large contributions from Zr$^0$ and
Zr$^+$ at the surface while the bulk is dominated by Zr$^{2+}$
contributions along with Zr$^0$ contributions. Zr$^+$ possessing
electronic configuration of [Kr]4$d^2$5$s^1$ is not a stable
configuration and hence is unstable in the bulk. Any such entity
would charge disproportionate to Zr$^0$ and Zr$^{2+}$.\cite{cmvarma}
The boron deficiency at the surface provides a reconstructed
electronic structure stabilizing the Zr$^+$ entities.

The features S1 and S2 (see Figs. 1 and 2) exhibit the energy
separation close to 2.3 eV, which is similar to the spin-orbit
splitting. These features could be fit with two peaks marked S1 and
S2 in Fig. 2 and are attributed to the satellite features associated
to the photoemission of 3$d$ electrons. The other possibility could
be loss features due to various collective excitations in the solid
such as plasmon excitations, phonon excitations etc. If that is the
case the signature of such loss features would appear with every
core level studied and we did not observe this to happen with any of
the boron core levels studied. Moreover, the energy separation of
such feature in Zr 3$p$ spectra shown in Fig. 3 is much larger than
3 eV observed in the 3$d$ core level spectra. Thus, one can rule out
the possibility of loss features in the present case. The intensity
of the satellite feature increases with the increase in bulk
sensitivity indicating that these are associated to the bulk
electronic structure dominated by Zr$^{2+}$ contributions.

The boron (B 1$s$) core level spectrum also shows an asymmetry
towards higher binding energies.\cite{Doniach} The binding energy of
B 1$s$ is consistent with the reported values in typical transition
metal borides\cite{Perkins} and borocarbides RNi$_{2}$B$_{2}$C (R =
Y and La)\cite{Fujimori} (187.1 - 188.3 eV). We have not observed
evidence of impurity phases such as B$_{2}$O$_{3}$ (B 1$s$; 191.10
eV) in our spectra in contrast to that found earlier in the case of
polycrystalline ZrB$_{12}$ sample.\cite{Flores} Although, all the
boron sites are equivalent from the crystal structure point of view,
observation of three distinct features with significantly different
binding energies indicate presence of more than one type of B in the
crystal. The intensity of the feature at 187 eV (feature D in Fig.
2) is more intense in the relatively more bulk sensitive spectra
indicating its bulk nature. These boron atoms presumably reside
closure to the Zr$^{2+}$ sites with largest effective negative
potential leading to a lower binding energy.

\begin{figure}
\vspace{-4ex}
 \includegraphics[scale=0.4]{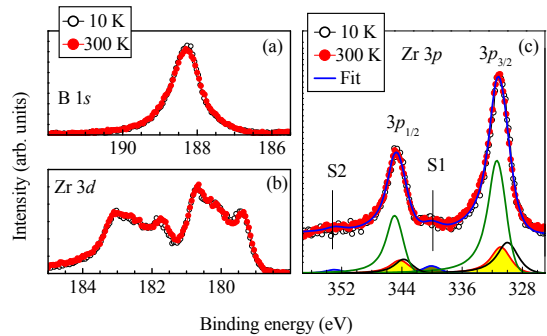}
 \vspace{-48ex}
 \caption{(a) B 1$s$, (b) Zr 3$d$ and (c) Zr 3$p$ core level spectra
at 10 K (open circles) and 300 K (closed circles). The lines in (c)
show the fitting results.}
 \vspace{-2ex}
\end{figure}

The spectra collected with $h\nu_1$ photon energy at 300 K and 10 K
are superimposed in Fig. 3 to see the effect of temperature.
Clearly, the line shape of all the core level spectra remain almost
identical at these two widely differed temperatures. Although 10 K
is little higher than the superconducting transition temperature,
the large change in temperature from 300 K to 10 K expected to
manifest the phonon induced effects. Almost identical shape and
relative intensity of the features in all the core level spectra
suggest that the thermal influence on the electronic structure, if
there is any, is below the detection level of the technique. In
order to investigate the signature of satellite and temperature
effects independently on other core level spectra, we analyze the Zr
3$p$ core level spectra in Fig. 3(c). Again, the spectral lineshape
is found to be identical at both the temperatures studied. The
spectra could be simulated using multiple Zr valencies along with
the signature of satellites S1 and S2 associated to the spin-orbit
split 3$p$ signals reestablishing independently the observations in
the Zr 3$d$ core level spectra.

The atomic number 40 of Zr corresponds to the electronic
configuration of [Kr]4$d^2$5$s^2$. Since, the B$_{12}$ unit has an
effective valency of (2-),\cite{Fillipov} the expected Zr valency in
this system would be Zr$^{2+}$ with an electronic configuration of
4$d^2$ - these two electrons will populate the $t_{2g}$ bands
leading to the metallic ground state. Since the satellite feature
represents the poorly screened feature in the photoemission, the
corresponding electronic state configuration will be $|4d^2>$ in the
photoemission final state. The main peak, C1 corresponds to the well
screened final state, where the positive charge due to the core hole
is screened by the transfer of an electron from the conduction band
and/or ligands leading to an electronic state $|4d^3\underbar{L}>$;
$\underbar{L}$ represent a hole in the ligand levels. The energy
separation, $\Delta E$ of the peak C1 to the satellite feature S1 is
about 3 eV. The intensity ratio of the peaks S1 and C1 is in the
range of 0.5 to 0.6, which is quite large considering the intensity
ratios found in various transition metal oxides such as
cuprates.\cite{dds} Considering the above discussed two final states
representing the electronic spectra within the first approximation,
the electron correlation strength is expected to be close to 3
eV.\cite{dds,corelevel}

%\section{Conclusions}

In summary, we studied the electronic structure of a BCS
superconductor, ZrB$_{12}$ employing high resolution photoemission
spectroscopy. Hard $x$-ray photoemission helped to reveal different
surface and bulk electronic structure of this compound. Experimental
results exhibit large Zr$^{2+}$ component in the bulk along with
some Zr$^{0}$ component. Multiple valencies of Zr appear at the
surface due to boron deficiency in the surface layers. In addition,
we observe signature of satellite feature in the Zr core level
spectra indicating finite electron correlation among conduction
electrons. Decrease in temperature down to 10 K does not have
significant influence in the spectral lineshape and/or energy
position indicating the influence from the lattice degrees of
freedom is below the detection limit. These results provide evidence
of different surface and bulk electronic structure even is a
conventional superconductor, ZrB$_{12}$ that is important to
consider while fabricating devices based on these materials.

%\section{acknowledgments}

The authors acknowledge financial support from the DST-DESY project
to perform the experiments at P09 beamline at PETRA III, Hamburg,
Germany and Dr. Indranil Sarkar for his help during the
measurements. The authors, K. M. and N. S. acknowledge the
Department of Science and Technology for financial assistance under
the Swarnajayanti Fellowship Programme.  G.B. wishes to acknowledge
financial support from EPSRC, UK (EP/I007210/1).


\begin{thebibliography}{}
%\bibitem{cab6}
% D. P. Young {\em et al.} Nature {\bf397}, 412 (1999);
%P. Vonlanthen {\em et al.} Phys. Rev. B {\bf62}, 10076 (2000); K.
%Maiti, V. R. R. Medicherla, S. Patil, and R. S. Singh, Phys. Rev.
%Letts {\bf99}, 266401 (2007); K. Maiti, Europhys. Letts {\bf82},
%67006 (2008).

\bibitem{surface}
K. Maiti, U. Manju, S. Ray, P. Mahadevan, I. H. Inoue, C. Carbone,
and D. D. Sarma, Phys. Rev.B {\bf73}, 052508 (2006); K. Maiti, A.
Kumar, D. D. Sarma, E. Weschke, and G. Kaindl, Phys. Rev. B {\bf
70}, 195112 (2004); K. Maiti, and  D. D. Sarma, Phys. Rev. B {\bf
61}, 2525 (2000); K. Maiti and  R. S. Singh, Phys. Rev. B {\bf 71},
161102(R) (2005). K. Maiti, R. S. Singh, and V. R. R. Medicherla,
Phys. Rev. B {\bf 76}, 165128 (2007).

\bibitem{Matthias}
B. T. Matthias, T. H. Geballe, K. Andres, E. Corenzwit, G. W. Hull,
and J. P. Maita, Science {\bf159}, 530 (1968).

\bibitem{Glaser}
B. Post and F. W. Glaser, J. Metals Trans. AIME {\bf631}, (1952).

\bibitem{Paderno}
Yu. B. Paderno, A. B. Liashchenko, V. B. Filippov, and  A. V.
Dukhnenko, {\it Advantages and  Challenges in Science for Materials
in the Frontier of Centuries}, edited by V. V. Skorokhod, (IPMS,
Kiev 347 2002).

\bibitem{Fisk}
Z. Fisk and B. T. Matthias, Science {\bf165}, 279 (1969).

%\bibitem{Alekseev}
%A. V. Rybina, K. S. Nemkovski, and P. A. Alekseev, Phys. Rev. B
%{\bf82}, 024302 (2010).

%\bibitem{Petrovic}
%J. Teyssier, R. Lortz, and  A. Petrovic, Phys. Rev. B {\bf78}, 134504 (2008).

%\bibitem{Wang}
%Y. Wang, R. Lortz, and Yu. B. Paderno, Phys. Rev. B {\bf72}, 024548
%(2005).

%\bibitem{Daghero}
%Daghero, R. S. Gonnelli, G. A. Ummarino, A. Calzolari, V.
%Dellarocca, V. A. Stepanov, V. B. Filippove, and Y. B. Paderno,
%Supercond. Sci. Technol. {\bf17}, S250 (2004).

%\bibitem{Gasparov}
%V. A. Gasparov, N. S. Sidorov, and I. I. Zver'kova, Phys. Rev. B
%{\bf73}, 094510 (2006).

\bibitem{Tsindlekht}
M. I. Tsindlekht, G. I. Leviev, I. Auslin, A. Sharoni, O. Millo, I.
Felner, Yu. B. Paderno, V. B. Filipov, and M. A. Belogolovski, Phys.
Rev. B {\bf69}, 212508 (2004).

\bibitem{Stout}
R. W. Mar and N. D. Stout, J. Chem. Phys. {\bf57}, 5342 (1972).

\bibitem{Bindu}
R. Bindu, G. Adhikary, N. Sahadev, N. P. Lalla, and K. Maiti, Phys.
Rev. B {\bf84}, 052407 (2011); V. R. R. Medicherla, S. Patil, R. S.
Singh, and Kalobaran Maiti, Appl. Phys. Lett. {\bf 90}, 062507
(2007).

\bibitem{Geetha}
G. Balakrishnan, M. R. Lees, and D. M. K. Paul, J. Crystal Growth
{\bf256}, 206 (2003).

%\bibitem{covalency}
%K. Maiti, Phys. Rev. B {\bf 73}, 235110 (2006); {\it ibid.}, Phys. Rev.
%B {\bf73}, 115119 (2006); {\it ibid.}, Phys. Rev. B. {\bf77}, 212407 (2008).

\bibitem{Lortz}
R. Lortz, Y. Wang, S. Abe, C. Meingast, Yu. b. Paderno, V. Filipov,
and A. Junod, Phys. Rev. B {\bf72}, 024547 (2005).

\bibitem{Doniach}
S. Doniach and M. S\'{u}njic, J. Phys. C: Solid State Phys. {\bf3},
285 (1970).

\bibitem{Flores}
L. Huerta, A. Duran, R. Falconi, M. Flores, and R. Escamilla,
Physica C {\bf470}, 456 (2010).

\bibitem{Morant}
C. Morant, J. M. Sanz, L. Galan, L. Soriano, and F. Rueda, Surface
Science {\bf218}, 331 (1989).

\bibitem{Wells}
A. F. Wells, {\it Structural Inorganic Chemistry}, edited by A.
F. Wells (Clarendon, Oxford, 1975).

\bibitem{Sarma}
L. Kumar, D. D. Sarma, and S. Krummacher, Appl. Surface Sci {\bf32},
309 (1988).

\bibitem{Takahashi}
F. Iga, Y. Takakuwa, T. Takahashi, M. Kasaya, T. Kasuya, and T.
Sagawa, Solid State Commun. {\bf50}, 903 (1984).

\bibitem{cmvarma} C. M. Varma, Phys. Rev. Lett. {\bf 61}, 2713 (1988).

\bibitem{Perkins}
C. L. Perkins, R. Singh,  M. Trenary, T. Tanaka, and Y. Paderno,
Surf. Sci. {\bf470}, 215 (2001); S. Patil {\it et al.}, Solid State
Commun. {\bf 151}, 326 (2011).

\bibitem{Fujimori}
K. Kobayashi, T. Mizokawa, K. Mamiya, A. Sekiyama, A. Fujimori, H.
Takagi, H. Eisaki, S. Uchida, R. J.  Cava, J. J. Krajewski, and W. F.
Jr. Peck, Phys. Rev. B {\bf 54}, 507 (1996).

\bibitem{Fillipov}
H. Werheit, V. Fillipov, K. Shirai, H. Dekura, N. Shitsevalova, U.
Schwarz, and M. Armbruster, J. Phys.:Condens. Matter {\bf23}, 065403
(2011).

\bibitem{dds} D. D. Sarma and S. G. Ovchinnikov, Phys. Rev. B {\bf
42}, 6817(R) (1990).

\bibitem{corelevel}
A. E. Bocquet, T. Mizokawa, K. Morikawa, A. Fujimori, S. R. Barman,
K. Maiti, D. D. Sarma, Y. Tokura, and M. Onoda, Phys. Rev. B {\bf
53}, 1161 (1996); K. Maiti, Priya Mahadevan, and D.D. Sarma, Phys
Rev B {\bf 59}, 12457 (1999).

%\bibitem{abhay}  A. N. Pasupathy ,  A. Pushp ,  K. K. Gomes , C. Parker ,
% G. Gu,  S. Ono,  Y. Ando, and  A. Yazdani, Science {\bf320}, 196 (2008).

\end{thebibliography}
\end{document}